\documentclass[fleqn,10pt]{wlscirep}
\usepackage{graphicx}
\begin{document}
\title{Band Structure of Topological Insulator BiSbTe$_{1.25}$Se$_{1.75}$}
\author[1,2]{H. Lohani}
\author[1]{P. Mishra}
\author[3]{A. Banerjee}
\author[3]{K. Majhi}
\author[3]{R Ganesan}
\author[4]{U. Manju }
\author[1,2]{D. Topwal}
\author[3]{P. S. Anil Kumar}
\author[1,2,*]{B.R. Sekhar}
\affil[1]{Institute of Physics, Sachivalaya Marg, Bhubaneswar 751005, India}
\affil[2]{Homi Bhabha National Institute, Training School Complex, Anushakti
Nagar, Mumbai 400085, India}
\affil[3]{Indian Institute of Science,  Bangalore- 560012, India}
\affil[4]{CSIR-Institute of Minerals and Materials Technology, Bhubaneswar- 751005, India}
\affil[*]{sekhar@iopb.res.in}



\begin{abstract}

We present our angle resolved photoelectron spectroscopy (ARPES) and density 
functional theory results on quaternary topological insulator (TI)
BiSbTe$_{1.25}$Se$_{1.75}$ (BSTS) confirming the non-trivial topology of the 
surface state bands (SSBs) in this compound. We find that the SSBs, which are 
are sensitive to the atomic composition of the terminating surface have a 
partial 3D character. Our detailed study of the band bending (BB) effects 
shows that in BSTS the Dirac point (DP) shifts by more than two times compared 
to that in Bi$_2$Se$_3$ to reach the saturation. The stronger BB in BSTS could 
be due to the difference in screening of the surface charges. From momentum
density curves (MDCs) of the ARPES data we obtained an energy dispersion relation showing the warping 
strength of the Fermi surface in BSTS to be intermediate between those found 
in Bi$_2$Se$_3$ and Bi$_2$Te$_3$ and also to be tunable by controlling the ratio
of chalcogen/pnictogen atoms. Our experiments also reveal that the nature of the BB effects are 
highly sensitive to the exposure of the fresh surface to various gas species. 
These findings  have important implications in the tuning of DP in TIs for 
technological applications.
\end{abstract}

\flushbottom
\maketitle
%
%
\thispagestyle{empty}


\section*{Introduction}

Discovery of the new quantum state of matter called topological 
insulators (TI) has attracted world wide interest due to their exotic 
properties which are manifestations of a non-trivial band 
topology\cite{rev,rev1}. TIs have insulating bulk and conducting edges due 
to the presence of some peculiar surface states (SSs). These SSs are spin 
non-degenerate with a unique property of spin momentum locking which results 
from the strong spin-orbit coupling (SOC) effects in combination with time 
reversal symmetry. It has been theoretically predicted that these SSs host 
many interesting properties like, Dirac fermion\cite{rev,dirac}, magnetic 
monopole\cite{mono} and Majorana bound state at the vortex in superconducting 
regime\cite{maj,maj1}. Strong immunity of these SSs to Anderson localization 
and backscattering in presence of non-magnetic impurities have tremendous 
technical advantages, especially for functional applications like spintronic 
devices and quantum computers\cite{tran,tran1}. Furthermore, tunability of the 
crossing point of the topological SSs, called the Dirac point (DP) by chemical 
doping, is another aspect important from such technological point of 
view\cite{levy,hatch,benia,dil}. In the known Bi and Sb based binary TIs the 
DP and the SSs are often obscured by contributions from bulk states. 
Tetradymite Bi$_2$Te$_2$Se which is isostructural to the prototypical TIs 
Bi$_2$Se$_3$ and Bi$_2$Te$_3$ has been found to be suitable for such tuning of 
the DP within the bulk band gap owing to its relatively large bulk 
resistivity\cite{rest1}. The resistivity can be optimized in the Sb doped 
quaternary alloy Bi$_{2-x}$Sb$_x$Te$_{3-y}$Se$_y$ by changing the ratio of the 
pnictogen (Bi and Sb) and chalcogen (Se and Te) atoms without disturbing its 
crystallinity. In this compound, topological nature with different bulk 
resistivity has been experimentally observed in a wide range of x and y 
combinations\cite{rest}. Thus, Bi$_{2-x}$Sb$_x$Te$_{3-y}$Se$_y$ provides an 
ideal platform to study the nature of topological surface states by tuning the 
Dirac node through controlling the proportion of chalcogen/pnictogen atoms.

Recently, quantum hall effect (QHE)\cite{tian} and scanning tunnelling 
spectroscopy (STS)\cite{ko} studies have been used to confirm the topological 
characters of BiSbTeSe$_2$ and Bi$_{1.5}$Sb$_{0.5}$Te$_{1.7}$Se$_{1.3}$. 
Tunability of the Dirac cone also has been observed in some of the 
compositions of Bi$_{2-x}$Sb$_x$Te$_{3-y}$Se$_y$ by using angle resolved 
photoelectron spectroscopy (ARPES) measurements\cite{sato,hor}.
Furthermore, the low bulk carrier density in these materials allows electrostatic
gating of the chemical potential letting a strong  control over electrical transport
properties which is vital for applications\cite{tian,iisc1,iisc2}.
 While most of the reported studies were focussed on the tunability by chemical doping or 
adding layers of other elements on the surface of TIs, the drifting of the 
topological surface state bands (SSBs) and DP with aging of the surface which 
are also important for device applications\cite{sensor}, has not been addressed 
adequately in this family of TIs\cite{levy,hatch}. In this paper, we present a 
detailed study of the electronic structure and aging effects of 
BiSbTe$_{1.25}$Se$_{1.75}$(BSTS) using ARPES in conjunction with density 
functional theory (DFT) based calculations. In the ARPES data, we observe 
topological character of the SSBs and a  warping of the Fermi surface 
(FS). These results are consistent with our calculated SSBs which fall within 
the region of bulk band gap of BSTS. In addition, we find pronounced effects of aging 
due to band bending (BB) and which are relatively stronger in this compound in 
comparison to  Bi$_2$Se$_3$. Effects of the BB  are enhanced due to the high 
adsorption of residual gases at low temperatures. Furthermore, experiments 
performed with constant dosing of different gases show that the BB effects are 
highly sensitive to the gas species.

\section*{Results and Discussion}

\begin{figure}
\begin{center}
\includegraphics[width=8cm,keepaspectratio]{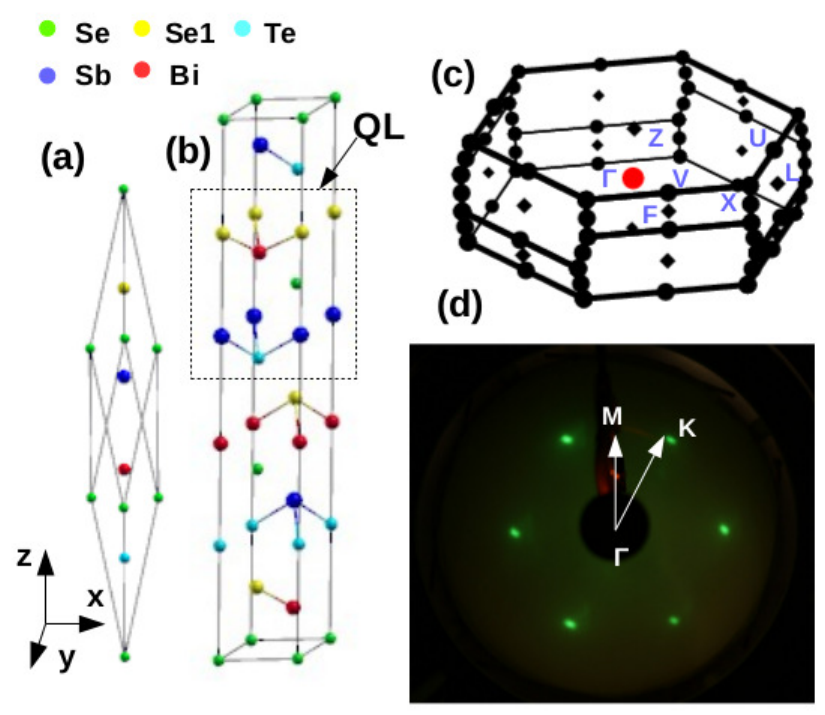}
\caption{\label{5crystal} (a) and (b) show the primitive and hexagonal unit 
cells of BiSbTeSe$_2$ respectively. The dotted box encloses the structure of 
five atomic layers which is called quintuple layer (QL). (c) Brillouin zone of 
the primitive unit cell where high symmetry k-points are marked. (e) LEED 
spots depict the hexagonal symmetry of the surface BZ of BSTS where the high 
symmetry k-directions $\Gamma$-K and $\Gamma$-M are marked.}
\end{center}
\end{figure}

Fig.\ref{5crystal}(a) shows the primitive unit cell of BiSbTeSe$_{2}$ which 
has a rhombhohdral symmetry. This structure can also be depicted as a 
hexagonal unit cell as shown in Fig.\ref{5crystal}(b). Basic building block of 
this structure is the so called quintuple layer (QL)  consisting of five 
atoms arranged in an order Se1-Bi-Se-Sb-Te. The Bi(Se1) atoms are connected to 
the Sb(Te) atoms with the center of inversion symmetry at the Se atom site. 
The structure of BiSbTeSe$_{2}$ is similar to that of Bi$_2$Se$_3$. 
Substitution sites of Sb and Te in Bi$_2$Se$_3$ to build the BiSbTeSe$_{2}$ 
structure for our calculations were chosen by considering the total energy minimization among 
various possible structures. Fig.\ref{5crystal}(c) shows the Brillouin zone 
(BZ) of the primitive unit cell where the high symmetry k-points are marked. 
In Fig.\ref{5crystal}(d) we have shown a low energy electron diffraction 
(LEED) pattern from our BSTS crystal depicting the hexagonal symmetry of the 
surface BZ. The BSTS crystal cleaves along the (111) crystal plane and 
scanning tunnelling microscopy experiments have shown that the terminated 
plane has Te/Se atoms on it\cite{ko}.

\begin{figure}
\begin{center}
\includegraphics[width=12cm,keepaspectratio]{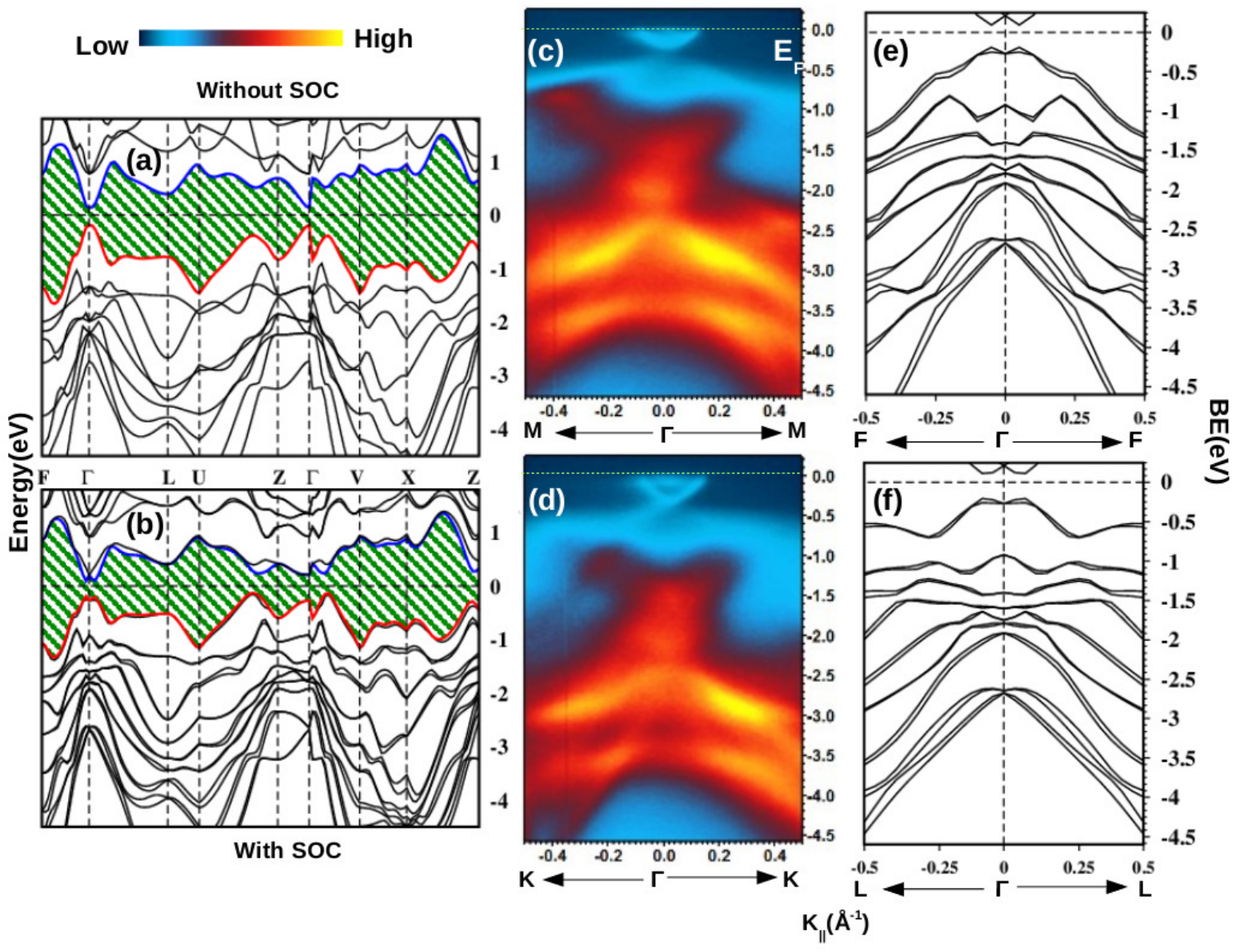}
\caption{\label{bstsband} (a) and (b) show the bulk band structure plots of 
BiSbTeSe$_{2}$ without and with inclusion of SOC effects respectively. (c) and 
(d) show the ARPES images of BSTS along the $\Gamma$-M and $\Gamma$-K 
directions of the surface BZ respectively. (e) and (f) show the bulk bands of 
(b) along the $\Gamma$-F and $\Gamma$-L directions respectively.}
\end{center}
\end{figure}

Fig.\ref{bstsband}(a) and (b) show the bulk band structure of nominal BiSbTeSe$_2$ 
composition without and with inclusion of SOC effects respectively. In both 
the cases, valence band(VB) and conduction band(CB) states are well separated 
in the energy scale. However, the SOC effects induce a small splitting in the bands 
along various k-directions. The band gap is $\sim$ 0.45eV at the $\Gamma$ point 
in SOC included case(Fig.\ref{bstsband}(b)) which is higher than the value 
0.3eV found in Bi$_2$Se$_3$\cite{bgap}. Structure of the top most VB(red) and 
the lowest CB(blue) along the F-$\Gamma$-L direction indicate a band inversion 
at the $\Gamma$ point after incorporating the SOC effects. This results in a 
non-trivial value of the Z$_2$ invariant in this system\cite{tang}. In 
Fig.\ref{bstsband}(c) and (d) the ARPES intensity plots of BSTS, which has a 
slightly different stoichiometry (BiSbTe$_{1.25}$Se$_{1.75}$) from the nominal 
composition (BiSbTeSe$_2$), are presented. These plots are taken along the 
$\Gamma$-M and $\Gamma$-K directions of the surface BZ by using 31eV photon 
energy respectively. Among the various bands seen in the VB region, the deeper 
lying bands with binding energy (BE) in the range E$_b$ = -0.5 to -4.5eV show 
a highly dispersive nature. A cone shaped distribution of low intensity is 
clearly visible in the vicinity of the E$_f$ around the $\Gamma$ point which 
is absent in the calculated band structure. The cone is formed by the 
topological SSBs in the bulk band gap region of the material. Although, 
inconsequential to the results of this study, a closer look at the data in the 
Fig.\ref{bstsband}(c) will reveal the existence a small energy gap near the 
tip of the cone which could be due to some possible misalignment of the 
angular position of the sample to the perfect $\Gamma$-M direction during the 
data collection. Nevertheless, in order to have a better comparison, raw data 
of the bulk bands along the $\Gamma$-F and $\Gamma$-L directions 
(Fig.\ref{bstsband}(b)) which correspond to the $\Gamma$-M and $\Gamma$-K 
directions of the surface BZ, are plotted adjacent to the ARPES images in 
Fig.\ref{bstsband}(e) and (f) respectively. Along both the directions,  
calculated bands placed at higher BE show a fair resemblance to the 
corresponding intensity pattern observed in the ARPES images.

\begin{figure}
\begin{center}
\includegraphics[width=12cm,keepaspectratio]{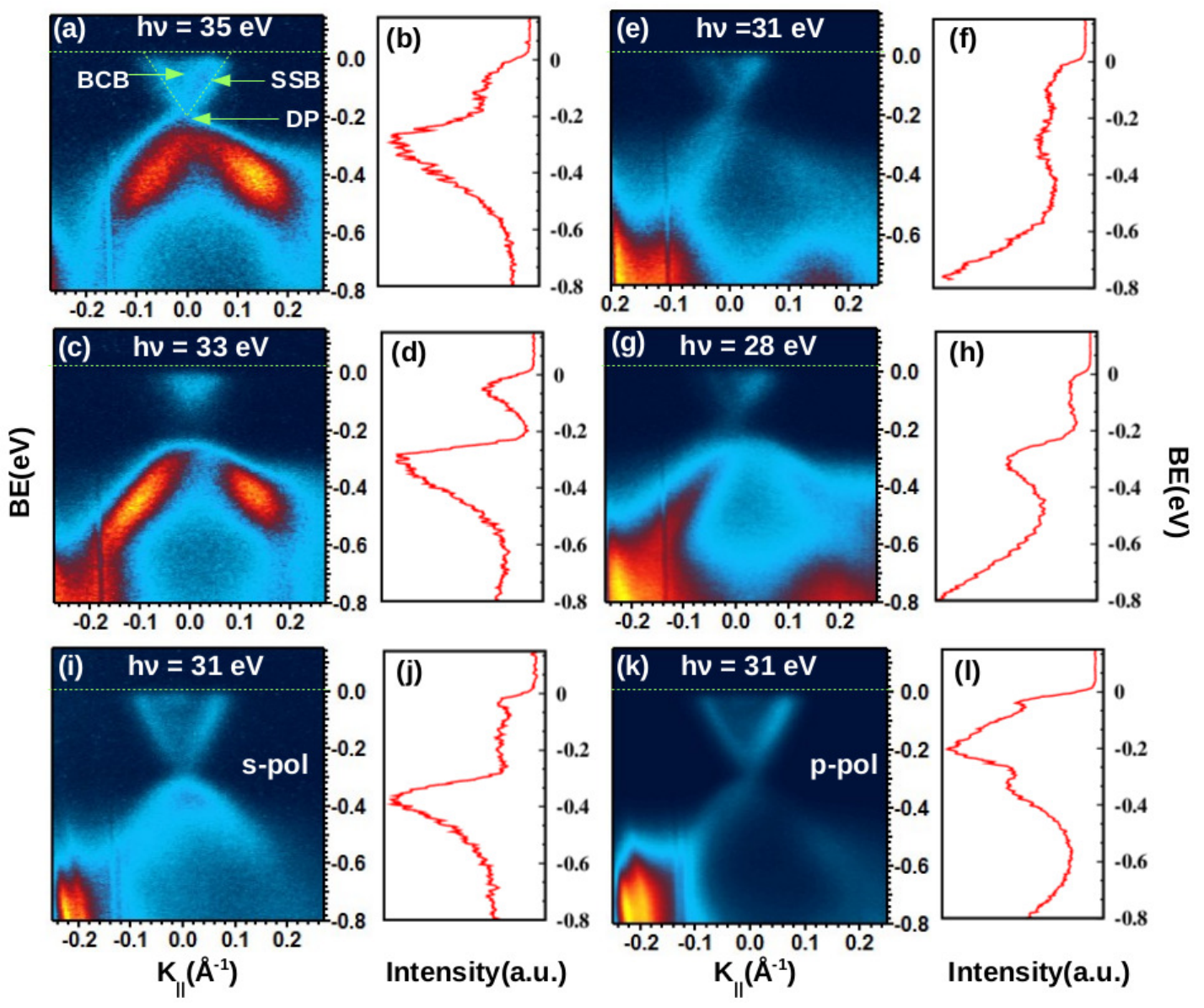}
\caption{\label{fs} (a), (c), (e) and (g) correspond to the near E$_f$ ARPES 
images taken at photon energy 35, 33, 31 and 28 eV respectively. (b), (d), (f) 
and (h) show EDC spectra corresponding to the images in (a), (c), (e) and (g) 
respectively. Intensity map using s and p-polarized photon energy of 31 eV are 
shown in (i) and (k) respectively. EDC spectra of these images are shown in 
(j) and (l) respectively.} 
\end{center}
\end{figure}

Fig.\ref{fs}(a) shows the near E$_f$ region of ARPES plot taken by using 35eV 
photon energy. The two SSBs are clearly visible exhibiting almost a linear 
dispersion. Intensity between these two SSBs indicates the presence of the bulk 
conduction band (BCB) states occupied due to n-type intrinsic impurities and 
defects. On the other hand, lower part of the Dirac cone strongly overlaps with
the bulk valence band (BVB) states which form a high intensity region at 
$\sim$ E$_b$ = -0.4eV. In order to identify the position of the DP, the 
corresponding energy density curve (EDC) of the image shown in Fig.\ref{fs}(a) 
is plotted in Fig.\ref{fs}(b). This EDC is taken at the $\Gamma$ 
point with k width of $\pm$ 0.02 \AA{}$^{-1}$. As is clear from this spectra 
the DP appearing at E$_d$ $\sim$ -0.2eV is obscured by the emission from the 
BVB states. In order to confirm the surface nature of the SSBs, ARPES images 
were collected at different photon energies as shown in Fig.\ref{fs}(a), (c), 
(e) and (g) which correspond to images taken at photon energy 35, 33, 31 
and 28 eV respectively. EDCs(at the $\Gamma$ point with k width of 
$\pm$ 0.02 \AA{}$^{-1}$) corresponding to these intensity plots are presented 
in Fig.\ref{fs}(b), (d), (f) and (h) respectively. The BVB gets modified 
sharply with the variation in photon energy indicating the bulk nature of 
these higher BE bands while the shape of near E$_f$ SSBs (upper part of Dirac 
cone) remains unaffected confirming the surface state character. The slight 
variation in the intensity of these SSBs is due to the difference in the 
matrix elements involved in the photoemission process\cite{dirac}. It should be 
noted that the SSBs show some significant changes close to the DP. The EDC 
spectra of 33eV photon energy (Fig.\ref{fs} (d)) shows an apparent opening of 
a gap in the SSBs in the vicinity of the DP (E$_d$ = - 0.2eV), unlike the 
case of 35eV photon energy (Fig.\ref{fs} (b)). Similarly, the spectral weight 
near $\sim$ -0.2eV BE in the EDC of 31eV (Fig.\ref{fs} (f)) also shows 
differences compared to that taken with 28eV (Fig.\ref{fs} (h)) photon energy. 
These results show that the SSBs are not of pure 2D character in this compound. 
The SSBs which mainly form the lower part of the Dirac cone hybridize with the 
BVB states and therefore acquire a partial 3D character. Origin of these 
hybridized states could be the impurities or defects in the system as 
suggested in a theoretical model for finite bulk band gaps by Black-Schaffer 
{\it et. al.} \cite{bal}. Experimental realization of such impurity induced 
gap opening in the SSBs at the DP has been recently reported by Sanchez-Barriga {\it 
et. al.} in their detailed ARPES study of (Bi$_{1-x}$Mn$_x$)$_2$Se$_3$ 
system\cite{rad}. Fig.\ref{fs}(i) and (k) show the ARPES plot taken with s and 
p-polarized light of photon energy 31eV respectively and adjacent 
Fig.\ref{fs}(j) and (l) display the EDC(at the $\Gamma$ point with k 
width of $\pm$ 0.02 \AA{}$^{-1}$) corresponding to them. In both the cases the 
linearly dispersive SSBs are clearly seen. However, the intensity of the BVBs 
at $\sim$ E$_b$ = -0.5eV got drastically reduced in the p-polarized case 
compared to the s-polarized. These changes are also visible in the spectral 
features of their EDCs (Fig.\ref{fs}(j) and (l)) showing the different orbital 
characters of these bands. 

\begin{figure}
\begin{center}
\includegraphics[width=8cm,keepaspectratio]{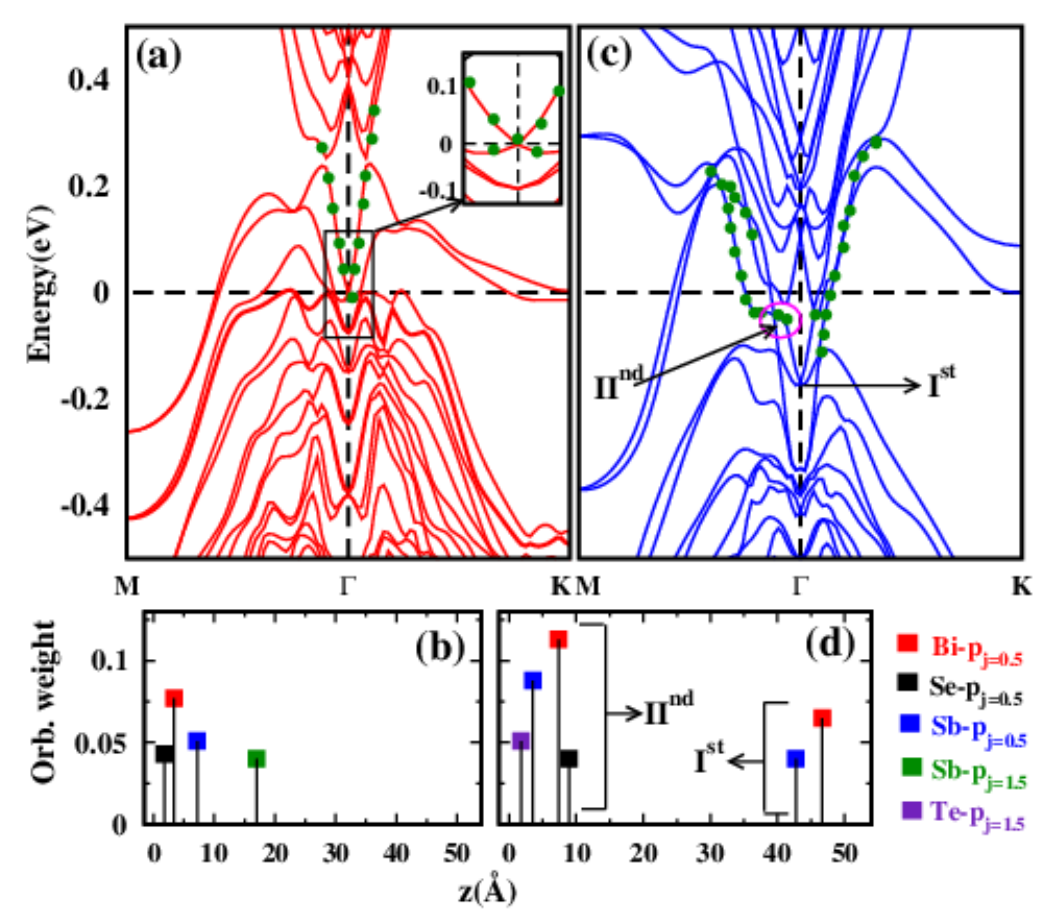}
\caption{\label{cal} (a) and (c) show calculated bands of Se1 and Te 
terminated face of 6QL slab geometry of BiSbTeSe$_{2}$ respectively. (b) 
Weight of atomic orbitals mainly contributing to the `V' shaped band (enclosed 
in the black box) at the $\Gamma$ point in Se terminated plain with respect to 
their distances from the surface. (d) Similar contribution to V shaped band at 
the $\Gamma$ point(I$^{st}$) and slightly away from the $\Gamma$ 
point(II$^{nd}$) in Te terminated face.} 
\end{center}
\end{figure}

Characteristics of the SSBs have been investigated by performing surface state 
calculations on the (111) crystal plane of BiSbTeSe$_{2}$. In Fig.\ref{cal}(a) 
bands (red) of 6QL slab structure with Se1 terminated face are plotted along 
the M-$\Gamma$-K direction of the surface BZ. It can be seen that two bands 
falling into the shape of `V' are observed just above the E$_f$ around the 
$\Gamma$ point. Here, green dots represent the orbital contribution coming 
from the atoms present in the topmost QL. Crossing of these Dirac like SSBs 
occurring at E$_f$ is more clearly visible in the inset. In order to figure 
out the origin of these `V' shaped bands, orbital projection weight of majorly 
contributing atoms to these bands at the $\Gamma$ point with respect to their 
distances from the surface (z) are plotted in Fig.\ref{cal}(b). 
Fig.\ref{cal}(b) shows clearly that the orbital character originates primarily
from the atoms close to the surface rather than the bulk region and the same 
character also persists in the nearby k-points of the $\Gamma$ point. This 
further confirms the surface state nature of these bands. These results show a 
qualitative similarity to the previously reported SSBs of 
BiSeTe$_2$\cite{tang}. The `V' shaped bands show a high resemblance to the 
intensity pattern of SSBs observed in the ARPES data (Fig.\ref{fs}(a)), 
though there is slight mismatch in the Fermi position. The mismatch is 
possibly due to the intrinsic n-doping in the sample which raises the E$_f$ 
level in the experimental data. The other possibility of Te terminated surface 
has also been examined and bands (blue) of 6QL slab of this geometry are shown in 
Fig.\ref{cal}(c). In this case, the `V' shaped band around the $\Gamma$ point (I$^{st}$ region) 
deviates from linearity as it moves away from the $\Gamma$ point (II$^{nd}$ 
region). Position of the tip of this V shaped band (180 meV) is quite
below the E$_f$, unlike the case of Se terminated face (Fig.4(a)). 
This energy position differs from the experimentally observed BVB (80 meV) of 
freshly cleaved BSTS (see Fig.1(e) of supplementary note).
 In addition, the orbital weight 
of these bands at region I$^{st}$ and II$^{nd}$ are dominated by atomic 
orbitals placed in the bulk and surface sites respectively as is clear from 
the Fig.\ref{cal}(d). Probably, this large mixing of bulk and surface 
characters leads to the deviation in the dispersion of this band. This 
different origin of the `V' shaped band around the $\Gamma$ point from   
bulk and surface in the Te and Se termination cases show that the SSBs are sensitive
to the atomic composition of the surface.

\begin{figure}
\begin{center}
\includegraphics[width=13cm,keepaspectratio]{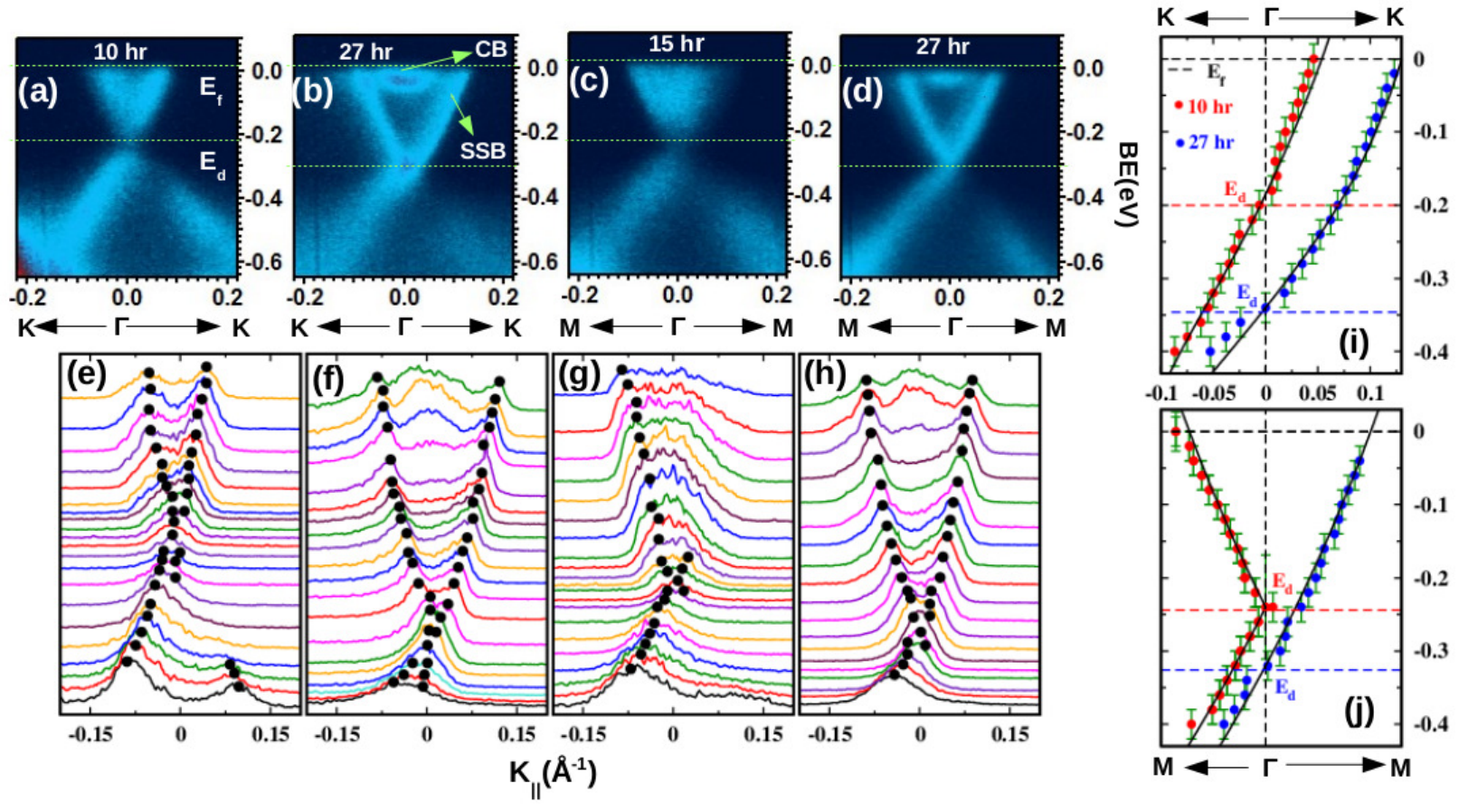}
\caption{\label{edcfit} (a) and (b) depict the ARPES intensity plots taken at 
31 eV photon energy along the $\Gamma$-K direction taken $\sim$ 10 and 27 hrs. 
after the sample cleaving. (c) and (d) show the same images along $\Gamma$-M 
direction collected at different time intervals from the cleaving. (e)-(h) 
show MDC spectra extracted from the images (a)-(d). The dispersion relation 
between E and k estimated from the MDC plots are fitted to the calculated 
values obtained from the model Hamiltonian\cite{wrap1} along the $\Gamma$-K 
(i) and the $\Gamma$-M (j) directions.}
\end{center}
\end{figure}

As mentioned before, the tunability of the DP within the bulk band gap of the 
system is an advantage of BSTS important from the technological point of view 
which could be achieved by chemical doping. Intimately related to this is the 
observed gradual shifting of the DP with adsorption of gases or even the 
elapse of time in ultra high vacuum after the crystal cleaving. This shifting 
of DP is caused by the band bending  which has been observed previously 
in various TIs, like Bi$_2$Se$_3$, Bi$_2$Te$_3$\cite{marco,qian}. We present 
our observations of BB effects on the SSBs in Fig.\ref{edcfit}, where 
\ref{edcfit}(a) and (b) show ARPES images along the $\Gamma$-K direction 
collected $\sim$ 10 and 27 hours after the cleaving. As can be seen from 
Fig.\ref{edcfit}(b), a significant shift of $\sim$ 0.14eV is observed in the 
position of E$_d$ in comparison to that in Fig.\ref{edcfit}(a). Further, the 
filled CB states in the nearby E$_f$ region around the $\Gamma$ point are 
clearly demarcated from the SSBs and an arc shaped structure is seen in these 
CB states which could be a signature of  two dimensional electron 
gas (2DEG) character arising due to the strong BB, like in 
Bi$_2$Se$_3$\cite{marco}. Similarly, shift of E$_d$ and appearance of distinct 
SSBs and CB states can also be found in the ARPES images Fig.\ref{edcfit}(c) 
and (d) which were collected along the $\Gamma$-M direction at different time 
intervals after the cleaving. Fig.\ref{edcfit}(e)-(h) show plots of momentum 
density curves (MDC) extracted from the ARPES images of Fig.\ref{edcfit}(a)-(d) 
respectively, where the linear dispersion of the MDC peaks is clearly seen. An 
energy dispersion relation of the SSBs can also be obtained from the model 
Hamiltonian approach proposed by Fu\cite{wrap1} by the following relation.

\begin{equation}
\begin{aligned}
&E_{\pm}(\vec{k}) = E_0(k) \pm \sqrt{\nu^{2}_k k^2 + \lambda^2 k^6 cos^2(3\theta)}\\
& \text{here}, \; E_0 = k^2/(2m^*)\quad;\quad\nu_k = \nu_0(1 + \alpha k^2)
\end{aligned}
\label{eq1}
\end{equation}
\noindent
where E$_{\pm}$ corresponds to the energy of the upper and lower band, E$_0$(k) 
generates particle-hole asymmetry, m$^*$ denotes effective mass, and $\theta$ 
indicates the azimuthal angle of momentum $\vec{k}$ with respect to the 
x-axis($\Gamma$-M direction). $\lambda$ is a parameter for the hexagonal 
warping. $\nu_0$ is the Dirac velocity which is modified to $\nu_k$ after including 
a second order correction parameter($\alpha$) to the Dirac velocity in the k.p 
Hamiltonian. The peak positions measured from the MDC plots along the 
$\Gamma$-K and $\Gamma$-M directions are fitted to the E-k dispersion relation 
of the SSBs obtained from the Eq.\ref{eq1} in Fig.\ref{edcfit}(i) and (j) 
respectively. The calculated bands nicely fit near the DP while a slight 
deviation can be seen in the regions away from the DP where the states of BVB 
and BCB are predominant. Parameters used for the fitting are tabulated in 
Table\ref{table} which shows that $\nu_0$ reduces significantly along the 
$\Gamma$-K direction after 27 hrs. from the cleaving in comparison to the 10 
hrs. cleaving case. On the other hand, warping strength, defined as 
$\sqrt{\lambda/\nu_0}$ remains almost constant under the influence of BB. 
 The estimated value of warping strength($\sqrt{\lambda/\nu_0}$ = 6.8) 
is intermediate between the value found in Bi$_2$Se$_3$ and 
Bi$_2$Te$_3$\cite{wrap2}.  
This result clearly establishes that FS warping and associated out of plane 
spin polarization can be controlled by the ratio of chalcogen/pnictogen atoms 
in Bi/Sb based TIs.

\begin{table}
\caption{\label{table} Parameters of calculated SSBs obtained after fitting 
with the E-k dispersion relation(Eq.\ref{eq1}) estimated from the ARPES data 
along the $\Gamma$-K and the $\Gamma$-M directions.}
\begin{center}
\begin{tabular}{ccccc}
\hline
&&$\Gamma$-K&&\\
\hline
\hline
Time(hr.)&1/2m$^\ast$(eV.\AA{}$^3$)&$\nu_0$(eV.\AA{})&$\alpha$(eV.\AA{}$^3$)&$\lambda$(eV.\AA{}$^3$)\\
\hline
\hline
10 &7&3.0&5&130\\
27 &1.8&1.8&5&80\\
\hline
&&$\Gamma$-M&&\\
\hline
\hline
15 &5&2.85&5&-\\
27 &4&2.65&5&-\\
\hline
\end{tabular}
\end{center}
\end{table}

\begin{figure}
\begin{center}
\includegraphics[width=11cm,keepaspectratio]{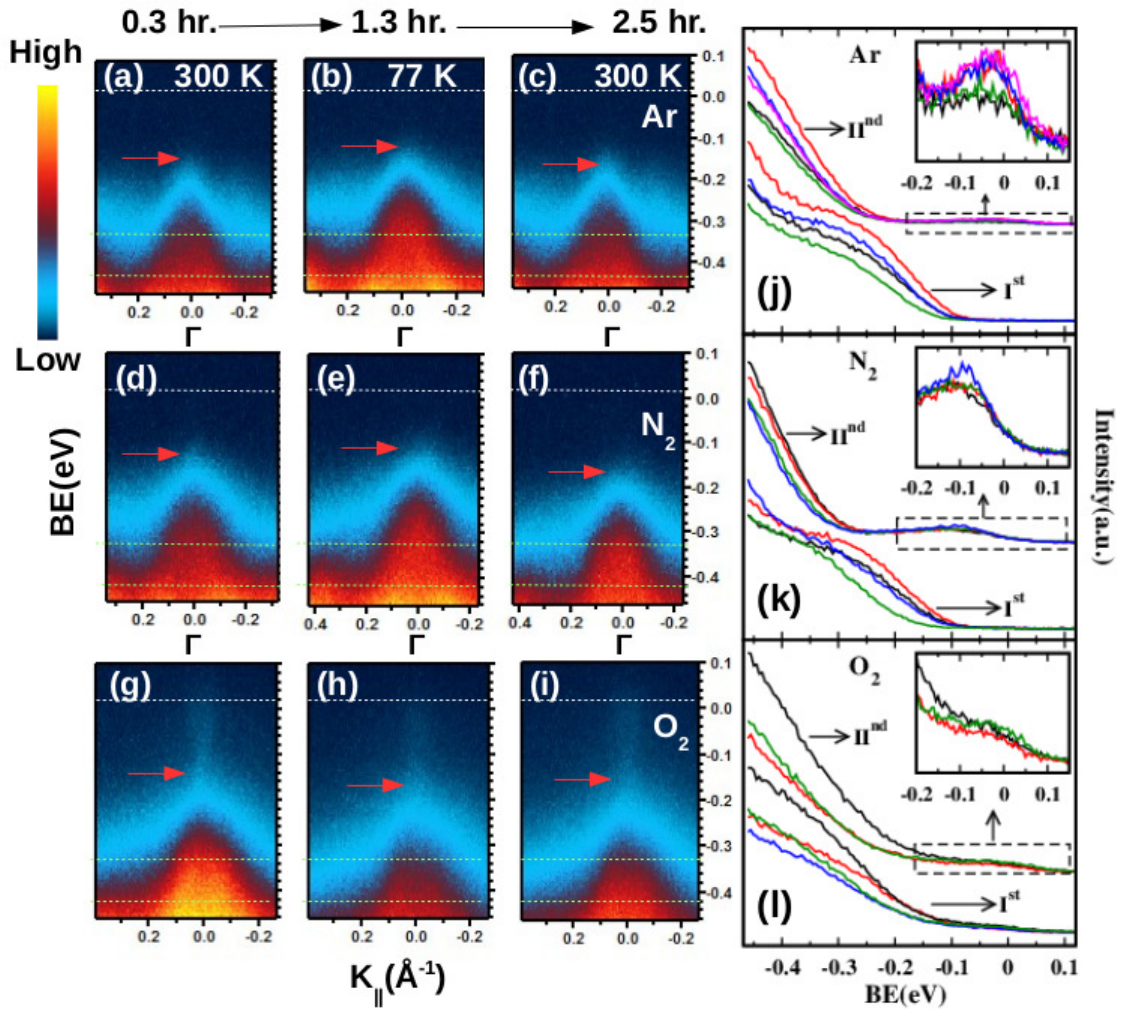}
\caption{\label{bbgas} (a), (b) and (c) show ARPES images taken at 300K - 77K - 
300K respectively just after the cleaving (I$^{st}$ thermal cycle) under 
constant Ar exposure. (d)-(f) and (g)-(i) display the ARPES images of the 
I$^{st}$ thermal cycle performed under constant dosing of N$_2$ and O$_2$ gases 
respectively. (j), (k) and (l) correspond to the EDC (taken around the 
$\Gamma$ point) plots of thermal cycling under the exposure of Ar, N$_2$ and 
O$_2$ gases respectively, where the inset of each plot shows the enlarged view 
of the near E$_f$ region. Different colours (black $\rightarrow$ red 
$\rightarrow$ green $\rightarrow$ blue $\rightarrow$ magenta) of the EDC 
spectra represent various stages (300K $\rightarrow$ 77K $\rightarrow$ 300K 
$\rightarrow$ 77K $\rightarrow$ 300K) of thermal cycling respectively.}
\end{center}
\end{figure}

Further experiments were performed to understand the BB effect by using a 
laboratory HeI (21.2 eV) photon source in combination with a Scienta R3000 
electron energy analyzer. It has earlier been proposed that the BB in TIs 
originate from accumulation of additional charges at the surface and these 
extra charges arise due to Se vacancies present in the bulk as well as those 
created at the surface in the process of surface cleaving\cite{BB1,BB2}. 
Moreover, adsorption of residual gases further changes the charge distribution 
at the surface\cite{hatch,benia,dil}. In order to understand the role of 
adatoms in BB, we undertook the ARPES measurements over a cycle of 
temperatures, 300K - 77K - 300K. Since, the adsorption of residual gases is 
faster at low temperatures the effect of BB is expected to enhance at low 
temperatures. This view is fairly supported by our thermal cycling ARPES data 
taken at different time intervals under  constant exposure of 
Ar, N$_2$ and O$_2$ gases. In first panel, Fig.\ref{bbgas}(a), (b) and (c) 
show the ARPES images taken at 300K - 77K - 300K respectively just after the 
cleaving (I$^{st}$ thermal cycle) in the Ar environment. Similarly, second 
(Fig.\ref{bbgas}(d)-(f)) and third (Fig.\ref{bbgas}(g)-(i)) panels display the 
ARPES images of the I$^{st}$ thermal cycle performed under constant dosing of 
N$_2$ and O$_2$ gases respectively. In Fig.\ref{bbgas}(b), a marginal shifting 
of the BVB maximum (marked with red arrow) is observed towards the E$_f$, 
though it was recorded at a later time in compared to Fig.\ref{bbgas}(a). 
This result of BB  is contrary to the behavior observed under the ultra 
high vacuum conditions. It indicates that the 
Ar adatoms act like an electron acceptor compensating the Se vacancy induced 
downward BB and thereby lead to a small upward shifting of the BVB. This 
inference is supported by the relatively large downward shift of the BVB 
due to the gas desorption in the next 300K annealed data (Fig.\ref{bbgas}(c)). 
These changes are more clearly visible in the higher BE region (between the 
two red dotted lines). Similar characteristics of hole doping are also seen 
under the exposure of N$_2$ gas (Fig.\ref{bbgas}(d)-(f)). However, the dosing 
of O$_2$ gas gives rise to an opposite BB effect {\it i. e.} features of 
n-doping as clear from Fig.\ref{bbgas}(g)-(i). Further in this case, a faint 
feature of the SSBs appears quite early at the top of the BVB, unlike the other gas 
exposure cases after the cleaving. 
This probably is linked to the higher adsorption of O$_2$ gas which 
accelerates the BB. These changes are also compared in the EDC (taken 
around the $\Gamma$ point of k width of $\pm$ 0.02 \AA{}$^{-1}$) plots 
Fig.\ref{bbgas}(j), (k) and (l) which correspond to the first, second and 
third panels respectively. Data sets of thermal cycling performed $\sim$ 0:30 
and 24:00 hrs. after the cleaving are marked I$^{st}$ and II$^{nd}$ 
respectively. In Fig.\ref{bbgas}(j), a sharp reduction is observed in the 
intensity of initial 300K spectra (black) in comparison to the 77K spectra 
(red) which again raise and fall in the next 77K (blue) and 300K (green) data. 
Similar behavior is reproduced at the II$^{nd}$ set of thermal cycling also. 
In addition, spectral weight originating from the filled BCB states also is 
found to be enhancing in the vicinity of the E$_f$ as shown in the inset. 
The trend under thermal cycling shown by the EDC spectra of N$_2$ case 
(Fig.\ref{bbgas}(k)) qualitatively matches with that of the Ar exposure case, 
whereas, the EDC plots of the O$_2$ exposure case (Fig.\ref{bbgas}(l)) show  
an opposite behavior. The small recovery of annealed 300K (green) spectra with 
respect to the 77K (red) in the II$^{nd}$ set of cycle could possibly be a 
signature of incomplete desorption of the O$_2$ gas.

It was reported that in the binary TI Bi$_2$Se$_3$ the DP moves by 116 meV 
from its initial position just after cleaving to a saturation value under the 
influence of BB\cite{levy}. Our own measurements also showed a 
shifting of $\sim$ 0.1eV in the time scale of 11:00 hr. after the sample cleaving 
in Bi$_2$Se$_3$. It is interesting to note that this movement is substantially 
lesser in comparison to that in BSTS where it is $\sim$ 0.2 eV in a similar time 
scale and experimental conditions (Fig.1 of supplementary note). Recent 
studies on Bi$_2$Se$_3$ have shown that not only the extra charges at the 
surface but also its periodic re-arrangement inside the bulk creates a 
Coulomb potential of long range order contributing to the BB\cite{BB2}. This unique property is inherent to the layered structure of 
Bi$_2$Se$_3$ where charge is accumulated and depleted at both ends of each 
QL. Introduction of additional elements (Sb and Te) in the QL of BSTS leads to 
an asymmetry in the structure of the QL compared to the Bi$_2$Se$_3$. The 
presence of Te atoms in addition to the Se atoms at the terminating faces of 
the QL  in Bi$_{1.5}$Sb$_{0.5}$Te$_{1.7}$Se$_{1.3}$ have been 
observed in STM measurements\cite{ko}. This could possibly provide different 
screening to the surface charges compared to the Bi$_2$Se$_3$. So an 
oscillatory behavior of charge density could persist at larger distances 
inside the bulk region and result in a Coulomb potential of higher and longer 
range giving rise to  stronger BB in BSTS in comparison to the 
Bi$_2$Se$_3$. Our experimental observations, stronger BB and lower DP position 
in BSTS compared to Bi$_2$Se$_3$, are consistent with the ARPES results on a 
similar composition Bi$_{1.5}$Sb$_{0.5}$Te$_{1.7}$Se$_{1.3}$ reported by 
Golden {\it et. al.} \cite{gol}. These authors have also attributed the difference in
effective screening of the adsorbate-induced surface charge to the variation in
the temporal evolution of the SSBs between Bi$_{1.5}$Sb$_{0.5}$Te$_{1.7}$Se$_{1.3}$ 
and Bi$_2$Se$_3$ compounds. In addition, they suggested that different compositions 
of the terminated face between the two compounds could also influence the 
sticking process of the residual gas atoms thereby leading to different BB 
behaviour. This argument is supported by our first principles results where we 
find that the nature of the SSBs are different in  Se and 
Te terminated slab geometries(Fig.\ref{cal}). Besides this, another factor affecting the BB 
process could be the difference in the relaxation process of the exposed 
surface of the two compounds. However, in case of Bi$_2$Se$_3$ Hofmann 
{\it et. al.} have ruled out any surface lattice relaxation from their LEED 
study\cite{marco}.

In conclusion, we discussed the results of our experimental studies using 
ARPES and first principles based Quantum Espresso band structure calculations 
and confirmed the non-trivial topology of the SSBs in BSTS. Our calculations 
show that the SSBs are sensitive to the atomic composition of the terminating 
surface and our experimental data shows that they have a partial 3D character. 
We have undertaken a detailed study of the shifting of the DP by the BB effect 
with elapse of time as well as adsorption of gases after the crystal cleaving. 
We find that under the BB effect the DP in BSTS shifts by more than two times 
compared to that in Bi$_2$Se$_3$ to reach a saturation. Our results suggest 
that the stronger BB in BSTS could be due to the difference in screening of 
the surface charges because of the different compositions of the QLs of the 
two compounds. From the MDCs of the ARPES data we obtained an energy dispersion 
relation showing the warping strength of the Fermi surface in BSTS to be 
intermediate between those found in Bi$_2$Se$_3$ and Bi$_2$Te$_3$ and also is 
tunable by the ratio of chalcogen/pnictogen atoms. Further experiments reveal that the 
nature of the BB effects are highly sensitive to the exposure of the fresh 
surface to various gas species; Ar and N$_2$ show signatures of hole doping 
while O$_2$ shows those of electron doping. Our findings could have importance 
in the tuning of the DP in topological insulators especially the members 
of the BSTS family for technological applications. 

\section*{Methods}

The high quality single crystal samples of BiSbTe$_{1.25}$Se$_{1.75}$(BSTS) 
used in this study were grown by modified Bridgman method. Stoichiometric 
amounts of Bi(99.999\%), Sb(99.999\%), Te(99.999\%) and Se(99.999\%) were 
heated in evacuated quartz ampoules to a temperature of 1073 K followed by 
slow cooling. Large sized single crystals($\sim$ 5 cm) were obtained which
cleaved easily along planes normal to the c-axis. The ARPES experiments were 
carried out using the facilities associated with the BaDELPH beamline of 
ELETTRA synchrotron center, Italy, equipped with a SPECS Phoibos 150 
hemispherical analyser. The photoemission spectra were collected on freshly 
cleaved ({\it in-situ} at 77K) surfaces of crystals under a vacuum of the 
order of 4.0 $\times$ 10$^{-11}$ mbar. In addition, ARPES data were taken by 
using our laboratory facility decked with a high flux GAMMADATA VUV He lamp 
(VUV5000) attached to a VUV monochromator (VUV5040) and a SCIENTA R3000 
analyser. Fermi energies of the samples were calibrated by using a freshly 
evaporated Ag film on the sample holder. The total energy resolution estimated 
from the width of the Fermi edge, was about 27meV for HeI excitation energy. 
The angular resolution was better than 1$^{\circ}$ in the wide-angle mode 
(8$^{\circ}$) of the analyzer. All the measurements were performed inside the 
analysis chamber under a base vacuum of 3.0 $\times$ 10$^{-10}$ mbar
 
First-principles calculations were performed by using a plain wave basis set 
inherent in Quantum Espresso (QE)\cite{qe}. Many electron exchange-correlation 
energy was approximated by the Perdew-Burke-Ernzerhof (PBE) 
functional\cite{Ernzerhof,Wang,Chevary}. Fully relativistic 
ultrasoft\cite{Vanderbilt} and non relativistic norm conserving 
pseudopotentials were employed for spin-orbit-coulpled (SOC) and non SOC 
calculations respectively. Fine mesh of k-points with Gaussian smearing of the 
order 0.0001 Ry was used for sampling the Brillouin zone integration, and 
kinetic energy and charge density cut-off were set to 100Ry and 450Ry 
respectively. Surface state calculations on (111) plane were performed by using 
supercell structures of hexagonal unit cell consisting of six quintuple layer (QL) 
with a vacuum separation of $\sim$ 26 \AA{}. Experimental lattice 
parameters\cite{rest} and atomic coordinates were relaxed under damped 
(Beeman) dynamics with respect to both ionic coordinates and the lattice 
vectors for all the structures.

\section*{Acknowledgments}
We acknowledge the scientific staff at the BaDELPH beamline of 
ELETTRA Sincrotrone Trieste, Italy for helping in the measurements.

\section*{Author contributions statement}
H. L., P. M., D. T.,   U. M. and  B. R. S. performed ARPES measurements.
A. B., K. M., R. G. and P. S. A. K. prepared the single single crystals and 
 characterized them. H. L. and B. R. S. analysed the ARPES data and wrote 
the manuscript. All authors discussed the results and commented on the
manuscript.

\section*{Additional information}

{\bf Supplementary information:} Supplementary information accompanies this paper at http://www.nature.com/srep
\vskip0.5cm
{\bf Competing financial interests:} We the authors declare no competing financial 
interests.
\vskip0.5cm
{\bf Data availability:} The datasets generated during and/or analysed during the current study are available from the corresponding author on reasonable request.

\end{document}